\documentclass[%
preprint,
superscriptaddress,
 amsmath,amssymb,
 aps, pre,
floatfix,
]{revtex4-2}

\usepackage{graphicx}
\usepackage{dcolumn}
\usepackage{bm}

\usepackage{enumitem}

\begin{document}

\preprint{APS/123-QED}


\title{Intrinsic Resonance depends on Network Size of Coupled-Delayed Interacting Oscillators}%

\author{Felipe A. Torres}
 \affiliation{Departamento de Computación e Industrias, Universidad Católica del Maule, Chile}
 \author{Alejandro Weinstein}
 \affiliation{Department of Electronics Engineering, Universidad Técnica Federico Santa María, Chile}
\author{Jesus M. Cortes}
\affiliation{Ikerbasque: The Basque Foundation for Science}
\affiliation{Bio-Bizkaia Health Research Institute, Spain}
\affiliation{Department of Cell Biology and Histology, University of the Basque Country, Spain}
\author{Wael El-Deredy}
\email{Contact author: wael.el-deredy@uv.cl}
\affiliation{Brain Dynamics Lab, Interdisciplinary Center of Biomedical and Engineering Research for Health, Universidad de Valparaíso, Chile}%

\date{\today}

\begin{abstract}
The collective frequency that emerges from synchronized neuronal populations—the network resonance—shows a systematic relationship with brain size: whole-brain's large networks oscillate slowly, whereas finer parcellations of fixed volume exhibit faster rhythms. This resonance–size scaling has been reported in delayed neural mass models and human neuroimaging, yet the physical mechanism remained unresolved.

Here we show that size-dependent resonance follows directly from propagation delays in delay-coupled phase oscillators. Starting from a Kuramoto model with heterogeneous delays, we linearize around the near-synchronous solution and obtain a closed-form approximation linking the resonance $\Omega$ to the mean delay and the effective coupling field. The analysis predicts a generic scaling law: $\Omega \approx (\sum_j c_{ij} \tau)^{-1}$, so resonance is delay-limited and therefore depends systematically on geometric size or parcellation density. We evaluate four growth scenarios—expanding geometry, fixed-volume parcellation, constant geometry, and an unphysical reference case—and show that only geometry-consistent scaling satisfies the analytical prediction. Numerical simulations with heterogeneous delays validate the law and quantify its error as a function of delay dispersion.

These results identify a minimal physical mechanism for size-dependent cortical resonance and provide an analytical framework that unifies numeric simulation outputs.
\end{abstract}

\maketitle


\section{\label{sec:Intro}Introduction}
Modeling natural systems as networks provides a powerful framework for understanding collective dynamics in complex systems, especially when oscillatory behavior and synchronization are involved. Network and graph theory offer quantitative tools for analyzing synchronization, stability, and frequency shifts across diverse domains such as biological, physical, and social systems \cite{boccaletti2006, Arenas2008}.

The Kuramoto model \cite{Kuramoto1975} stands as a foundational contribution to collective dynamics, demonstrating how a set of coupled oscillators can synchronize their phases and frequencies when the coupling strength surpasses a critical threshold. This gives rise to phase locking and coherent patterns of activity. Strogatz \cite{Strogatz2000} expanded this framework by analyzing how network topology, intrinsic oscillator properties, and coupling strength jointly determine synchronization dynamics, including transitions from partial to full synchronization.

However, real-world networks often include transmission delays, which introduce new layers of complexity. These delays shift phase relationships, alter the temporal structure of interactions, and often lead to metastability---where synchronization is transient and localized. Niebur \cite{Niebur1991} showed that delays systematically reduce the collective frequency—the emergent frequency at which a group of synchronized oscillators operate—relative to the average of their intrinsic frequencies. This phenomenon echoes the concept of resonance in dynamical systems, even in the absence of external stimulation, where emergent frequencies arise from the interplay of intrinsic oscillator properties, network topology, coupling strength, and temporal delays.

These principles have found direct application in modeling large-scale brain network dynamics. Kuramoto-like oscillators have been used to construct biologically plausible whole-brain network models constrained by empirical structural connectivity from the human connectome \cite{Cabral2014, Breakspear2010, Torres2024}. In such models, oscillators represent brain regions (typically with $\sim$40 Hz intrinsic frequency), and the coupling weights and delays reflect empirical inter-node connections. These interactions generate metastable states and broad frequency spectra, mimicking dynamic patterns seen in neuroimaging studies. Notably, transiently synchronized subnetworks have been shown to exhibit an inverse relationship between collective frequency and network size \cite{LeaCarnall2016, Torres2024}, suggesting that the scale of a network fundamentally shapes its emergent dynamics. 

 Further, human brain dynamics measured by electromagnetic mechanisms have geometric constraints. It is possible to obtain a better aproximation of cortical dynamics with a linear combination of geometrical modes of the brain rather than using a linear combination of the modes of the topology of the adjacency network \cite{Pang2023}. The need of spatial constraints of the complex networks was also shown as relevant to obtain the collective behaviour in other species' brains \cite{oDea2013, salova2025}. Then, the analysis of the geometry or spatial embedding of the networks could offer a deeper understanding about the emergent collective phenomena than relaying solely on network topology.

Despite its fundamental implications, the impact of network size on collective frequency has not been systematically analyzed. Most previous studies normalize the input to each node relative to the total number of nodes \cite{Strogatz2000, Lee2009}, which maintains local balance but potentially obscures global dynamical effects. In analogy to physical systems, increasing network size can lead to either denser packing (akin to higher pressure at fixed volume) or spatial expansion (greater distances at constant interaction strength). Both scenarios suggest a reduction in collective frequency when increasing size due to greater interaction complexity and longer effective delays.

Moreover, the changes in the connectivity density of the networks is known as an issue in the comparison between networks of different number of nodes by the varying of the the graph metrics. To overcome this difficulty, it was shown by \cite{vanWijk2010} that keeping the ratio between the degree and the number of nodes also keeps around similar values of the path length and clustering coefficient for lattice, small-world and random networks. In this sense, the fully-connected networks keeps the maximal ratio of the degree over the number of nodes. However, brain-like networks are sparse. It is expected that the connectivity density remains constant or even decrease by increasing the number of nodes in the fixed volume of an human brain. Connectivity density is also expected to be similar or decrease preserving the volume that a node represents in the brains of species of different sizes.\cite{salova2025}.

Here we examine steady-state response in the frequency behavior of Kuramoto-like networks in the presence of delays, similar to \cite{Cabral2014}. We analytically derive the expression of the collective frequency as a function of the number of nodes in the network. We define four distinct structural growth scenarios to characterize different ways of increasing network size. Case I (No geometry or delay scaling) serves as a conceptual baseline with fixed or absent delays \cite{Albert2002}. Case II (Growing weights with fixed delays) explores stronger coupling without spatial expansion (as increasing activity in a social network \cite{iniguez2023universal}), but is physically unrealistic considering geometric constraints. In contrast, Case III (Expanding the volume with spatial scaling) \cite{LeaCarnall2016} and Case IV (Increasing density in fixed volume) model biologically and physically plausible growth \cite{Daqing2011}. We also show practical examples of Case III building a network with its nodes located at a circumference with varying radius (Case A) and a practical example of case IV, increasing the number of nodes without changing the radius (Case B).     
For all cases, we identify the conditions under which network resonance emerges---marked by an emergent collective frequency that scales inversely with the number of nodes. Our findings lay the foundation for a deeper understanding of synchronization in large-scale networks, with implications for neuroscience, control theory, and the study of emergent behaviors in complex systems.

\section{Methods}

We define a fully connected, undirected graph $G=\{V, E\}$, where each pair of nodes $i, j$ is linked symmetrically with connection weights $c_{ij} = c_{ji}$, forming the connectivity (adjacency) matrix $\mathbf{C}$, and  $\tau_{ij}=\tau_{ji}$ forming the matrix of connection delays  $\boldsymbol{\tau}$. We assume $c_{ij} \geq 0$ and exclude self-connections, i.e., $c_{ii} = 0$, so $\mathbf{C}$ is symmetric and nonnegative.

On top of this graph, we insert a dynamical model consisting of 
$N$
 coupled Kuramoto-like oscillators with interaction delays \cite{Cabral2014}, i.e., 
\begin{equation}
    \dot{\theta}_i = \omega_i + \frac{K}{N} \sum_{j=1}^{N} c_{ij} \sin\left[\theta_j(t - \tau_{ij}) - \theta_i(t)\right],
    \label{eq:delayed_kuramoto}
\end{equation}
where $\theta_i(t)$ is the phase of the $i$-th oscillator, $\omega_i$ its natural frequency, $i = 1, \ldots, N$, $c_{ij}$ the weight of the connection from $j$ to $i$, $\tau_{ij}$ the delay, and $K$ the global coupling strength.

For analytical tractability, we assume that the network operates  in a near-synchrony regime, i.e. $|\theta_j(t - \tau_{ij}) - \theta_i(t)+2n\pi| \ll 1,\ \{n \in N_{0}\}$, allowing us to  linearize the sine function. Additionally, we approximate the delayed phase as a linear function of time, $\theta_j(t - \tau_{ij}) \approx \theta_j(t) - \Omega_j \tau_{ij}$, where $\Omega_j$ denotes the frequency of oscillator $j$.  Under these assumptions, the dynamics reduce to the following linearized system:

\begin{align}
    \dot{\theta}_i &\approx \omega_i - \frac{K}{N} \sum_{j=1}^{N} \left[ c_{ij}(\theta_i(t)-\theta_j(t)) + c_{ij} \Omega_j \tau_{ij} \right].
\end{align}

\subsection{Network size dependence of collective frequency}
Letting $\boldsymbol{\theta} = [\theta_1, \dots, \theta_N]^T$, $\boldsymbol{\omega} = [\omega_1, \dots, \omega_N]^T$, and $\boldsymbol{\Omega} = [\Omega_1, \dots, \Omega_N]^T$, the linearized dynamical system can be expressed compactly as:
\begin{equation}
    \dot{\boldsymbol{\theta}} = \boldsymbol{\omega} - \frac{K}{N} \left[ \mathbf{L} \boldsymbol{\theta} + (\mathbf{C} \odot \boldsymbol{\tau}) \boldsymbol{\Omega} \right],
    \label{eq:general_vector}
\end{equation}
where $\mathbf{L}$ is the graph Laplacian, and  $\odot$ denotes element-wise product. At the limit-cycle steady state, where 
$\dot{\theta}_i = \Omega_i$, the phase evolves linearly as $\theta_i=\Omega_i t$. Substituting into Eq.~\eqref{eq:general_vector}, we obtain:
\begin{equation}
    \left(\mathbf{I}+\frac{K}{N}\left[t\mathbf{L}+ (\mathbf{C} \odot \boldsymbol{\tau})\right]\right)\boldsymbol{\Omega}=\boldsymbol{\omega},
    \label{eq:hetero_delay_solution}
\end{equation}
and by Laplace transforming   Eq.~\eqref{eq:hetero_delay_solution}, it yields 

\begin{equation}
    \boldsymbol{\Omega}(s) = \left( s\mathbf{I} + \frac{K}{N} [\mathbf{L} + s(\mathbf{C} \odot \boldsymbol{\tau})] \right)^{-1} \boldsymbol{\omega}(s)s^2
    \label{eq:hetero_delay_solution_laplace}
\end{equation}
providing an explicit dependence on network size for the limit-cycle system response $\boldsymbol{\Omega}$. Under the near-synchrony assumption, all oscillators share a common frequency, i.e.,  $\Omega_i = \Omega$ for all $i$, where $\Omega$ denotes the collective frequency of intrinsic resonance.

\subsection{Homogeneous delays}
If $\tau_{ij} = \tau$, then $(\mathbf{C} \odot \boldsymbol{\tau}) \boldsymbol{\Omega} = \tau \mathbf{C} \boldsymbol{\Omega}$. 
Substituting it in Eq.~\eqref{eq:hetero_delay_solution}, and averaging over the nodes both sides in Eq.~\eqref{eq:hetero_delay_solution}, it yields  

\begin{align}
    \Omega &= \frac{\langle \omega \rangle}{1 + \frac{K\tau M}{N}},
    \label{eq:sync_omega}
\end{align}
where $M = \sum_j c$ is the row sum of $\mathbf{C}$. The average over the nodes of $\mathbf{L}$ for the defined $\mathbf{C}$ matrix is zero, removing the temporal dependence. As a control, if the homogeneous weight is $c_{ij}=c=1, \forall\{i,j\ |\ j\neq i\}$, then $M=N-1$, that in the limit $N \to \infty$, $M \approx N$, reducing  to the Niebur formula \cite{Niebur1991, Lee2009}:
\begin{equation}
    \Omega = \frac{\langle \omega \rangle}{1 + K\tau}.
\end{equation}

\subsection{Heterogeneous delays}
If $\tau_{ij}$ has distinct values for each edge of the network and $\boldsymbol{\omega}(s)=\frac{\boldsymbol{\omega}}{s}$ we obtain the steady state value of $\boldsymbol{\Omega}$ by the final time theorem
\begin{equation}
\boldsymbol{\Omega}=\lim_{s\rightarrow 0} s\boldsymbol{\Omega}(s)=\left(\mathbf{I}+\frac{K}{N}\mathbf{C}\odot\boldsymbol{\tau}\right)^{-1} \boldsymbol{\omega}.
\label{eq:Vector_ss_frequency}
\end{equation}
Then, the collective frequency is $\Omega=\langle \boldsymbol{\Omega}\rangle$.

\subsection{Numerical simulations}
Fully connected networks with varying number of nodes  from  $N=5$ to $N=100$ were simulated using the KuramotoNetworksPackage \cite{knp} for 10 seconds, and for ten initial conditions for the intrinsic frequencies and initial phases. 
The connectivity and delay matrices $\mathbf{C}$ and $\boldsymbol{\tau}$ have homogeneous off-diagonal elements for cases I-IV and heterogeneous values for cases A-B. In order to get heterogeneous weight values for the case A, we use the Gaussian filtered distance of the edge
\begin{equation}
    c_{ij}=c_o\exp\left(\frac{-\left(\frac{\tau_{ij}}{\tau_0}\right)^2}{\sqrt{2\pi}}\right),
    \label{eq:heterogenous_weights}
\end{equation}
which provides a isotropic connectivity that is a common practice in absence of detailed anatomical data \cite{Coombes2010}.
For all cases, $\mathbf{C}$ and $\boldsymbol{\tau}$ matrices have zeros in the main diagonal. The intrinsic frequencies $\omega_i$ of the nodes followed a Gaussian distribution with a 40 Hz mean and a standard deviation of 2 Hz for all the cases. The parameters $c_0,\ \tau_0$, and $K$ were selected to operate in a near-synchrony regime for almost any number of nodes in each case. The initial guess for the value of $K$ comes from $K_c>2\sigma\sqrt{\frac{2}{\pi}}$ for Gaussian distribution of the intrinsic frequencies \cite{Oliveira2020}, where $\sigma$ is the standard deviation in radians/seconds. 

We analyzed the steady state response of the simulations. The first second was discarded to avoid the transient dynamics starting from random initial phases. The steady-state frequency of each node is the peak frequency of Welch's periodogram using Hann time windows of 5 s with 50\% overlap. The collective frequency is taken to be the average of the frequencies from the $N$ nodes. 
Network synchrony was assessed by the temporal average of the Kuramoto Order Parameter (KOP) defined as 
\begin{equation}
    \operatorname{KOP}(t)=\left|\sum_{n=1}^{N}\exp(\mathrm{j}\theta(t))\right|
\end{equation}

\section{Results}

\begin{figure*}[!ht]
    \centering
    \includegraphics[width=\linewidth]{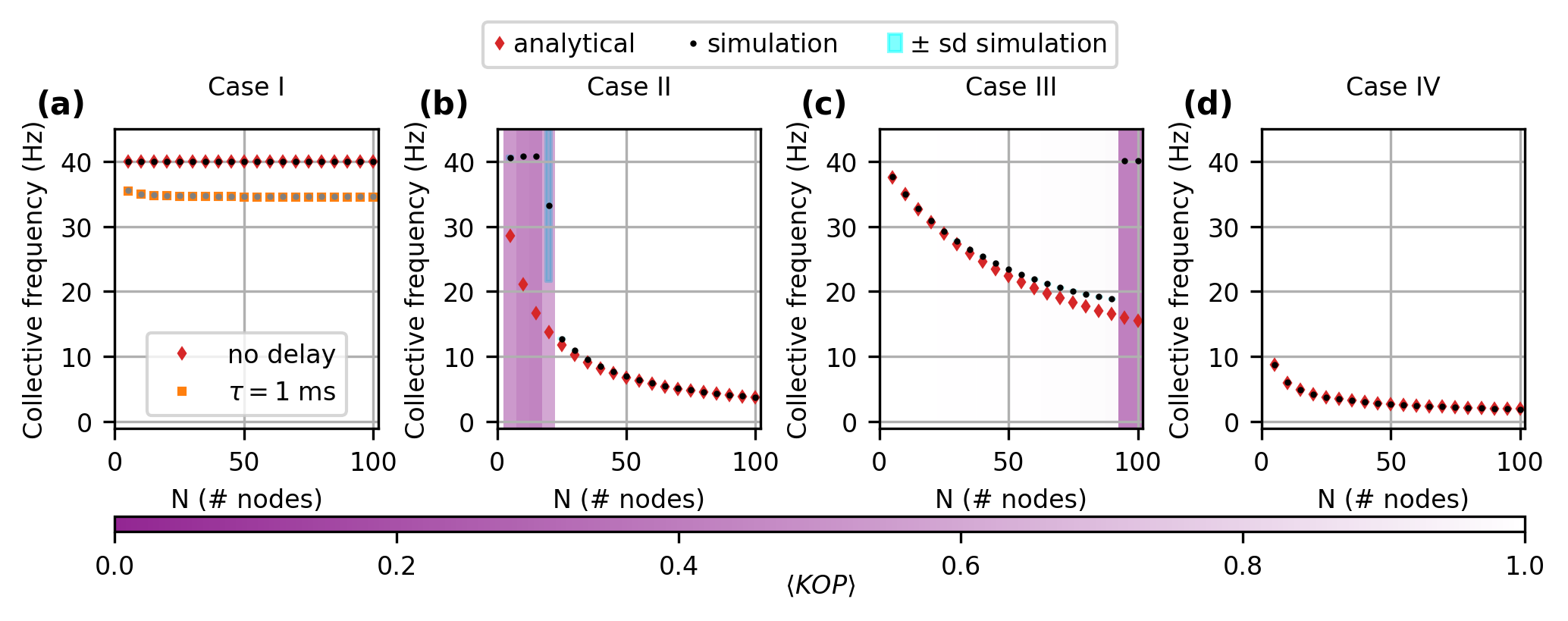}
    \caption{{\bf Dependence of Collective Synchronization Frequency on Network Size. Homogeneous connections.}
    (a) Case I: Homogeneous weights and no spatial embedding. Collective frequency remains equal to the mean intrinsic frequency ($\langle \omega \rangle = 2\pi \cdot 40$ Hz, SD = $4\pi$) for all $N$ without delays; with fixed homogeneous delays ($\tau=1$ ms, $K=160$), $\Omega$ converges to the Niebur limit \cite{Niebur1991} as $N\rightarrow\infty$. 
    (b) Case II: Weight scales with $N$ ($c=N$), but delays remain fixed ($\tau=10$ ms). Analytical and numerical frequencies decrease with $N$; mismatch for small $N$ arises from desynchronization (KOP$< 0.2$, shadowed region, $K=10$).
    (c) Case III: Spatial expansion, delays grow with $N$ while weights stay fixed. Analytical and numerical results match across $N$ ($\tau_0=0.1$ ms, $K=160$). Collective frequency drops with $N$; synchrony deteriorates beyond critical size (KOP $\approx 0$, shadowed region).
    (d) Case IV: Increasing network density in fixed volume, with $c = c_0/\tau^2$ and $\tau = \tau_0/\sqrt{N}$. synchrony sustained (KOP $\approx 1$) under $K=10$, $c_0 = 10^{-4}$, $\tau_0 = 5$ ms.}
    \label{fig:fig1}
\end{figure*}

\begin{figure*}[!ht]
    \centering
    \includegraphics[width=\linewidth]{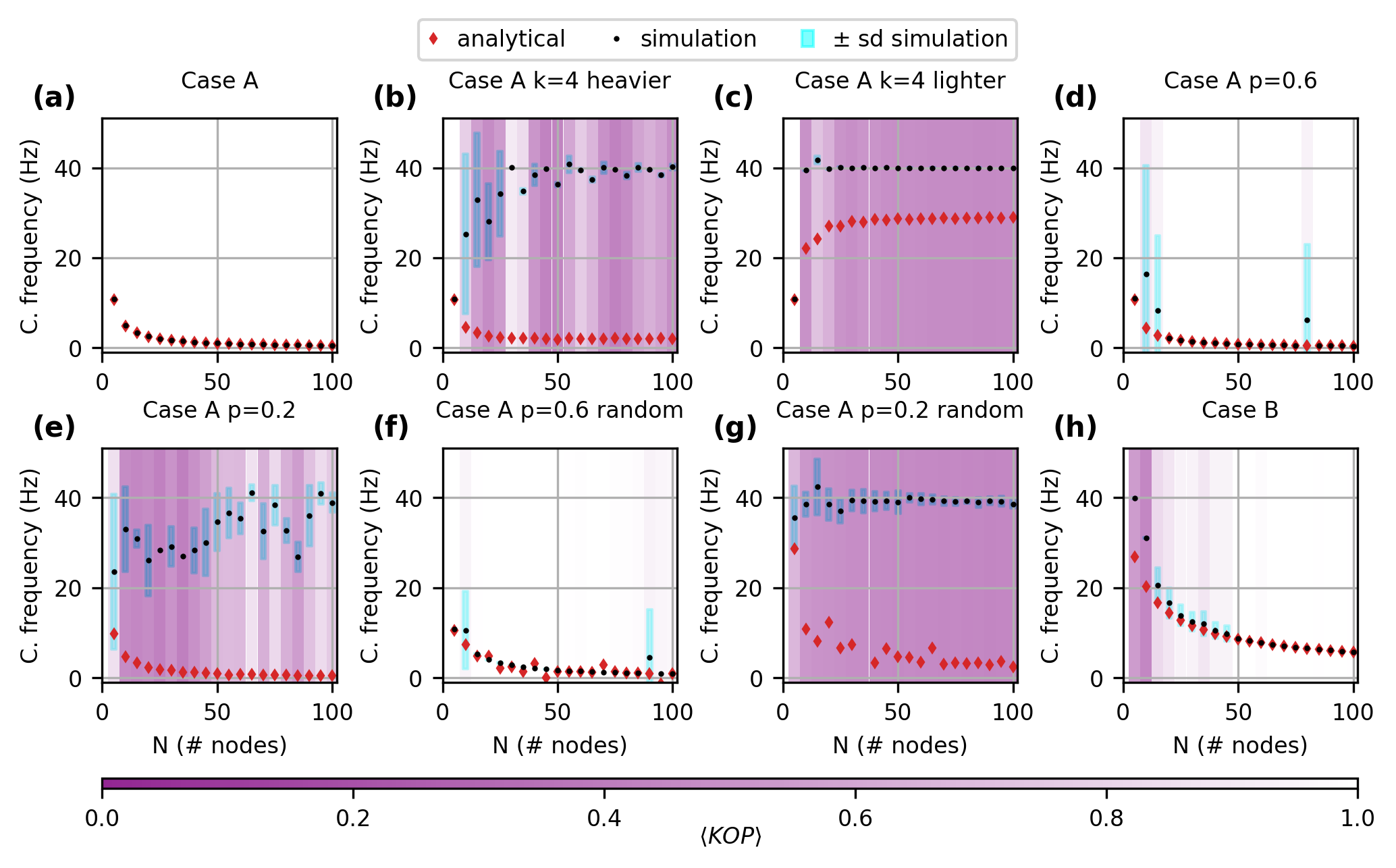}
    \caption{{\bf Dependence of Collective Synchronization Frequency on Network Size. Heterogeneous Connections.}
    (a) Case A: Distances scale along with the number of nodes in a fully connected network ($K=10000$, $\tau_0=1$ ms, $c_0=1$). The numerical results follow the analytical prediction of decreasing collective frequency as the network size increases.
    (b) Case A with fixed degree $k=4$ using the nearer nodes, which are the heavier connections. The collective frequency is lower than the intrinsic but does not follow the analytical prediction with low synchrony. 
    (c) Case A with fixed degree $k=4$ using the farther nodes, which are the lighter connections. For any $N>5$ the collective frequency remains near the average intrinsic frequency.
    (d) Case A with fixed connectivity density $p=0.6$ connecting the nearer nodes. The collective frequency follows the analytical prediction if the network synchronizes.
    (e) Case A with fixed connectivity density $p=0.2$ connecting the nearer nodes. The collective frequency is lower than the intrinsic average, but does not follow the analytical prediction as the networks' synchrony is low.
    (f) Case A with fixed connectivity density $p=0.6$ with random connections. The analytical and numerical results are close.
    (g) Case A with fixed connectivity density $p=0.2$ with random connections. The networks do not achieve synchrony and the collective frequency remains near the average intrinsic frequency.
    (h) Case B: Increasing the number of nodes in a fixed spatial area ($K=0.1$, $\tau_0=25$ ms, $c_0=1$). The numerical and analytic results are close. The collective frequency show more variance for the number of nodes where the KOP is high but not equal to 1.
    }
    \label{fig:fig2}
\end{figure*}

We analytically and numerically assessed how network size affects the collective frequency $\Omega$ under four archetypes of structural growth using homogeneous connection weights and delays (Fig. \ref{fig:fig1}). We also present the numerical results from two cases using heterogeneous connection weights and delays. In these last two cases, we used a growing spatial circular network with its nodes located in the circumference (Fig. \ref{fig:fig2}):

\paragraph{Case I: No geometry or delay scaling.}
In the absence of delays, Eq.~\eqref{eq:sync_omega} simplifies to $\Omega = \langle \omega \rangle$ for any $N$, confirming Strogatz’s classic result \cite{Strogatz2000}.  $\Omega = \langle \omega \rangle / (1 + MK\tau/N)$,  for a homogeneous delay $\tau>0$, which it also converges to the Niebur limit \cite{Niebur1991} for large $N$. Simulation and analytic results perfectly match (Fig.~\ref{fig:fig1}a).

\paragraph{Case II: Growing weights with fixed delay.}
Setting $c = N$, $\tau = \tau_0$, and $M = N(N-1)$, the analytical expression becomes $\Omega = \langle \omega \rangle / (1 + (N-1)K\tau_0)$. Synchrony fails for small $N$ ($\langle KOP\rangle<1$), leading to deviation from theory (Fig.~\ref{fig:fig1}b) as the near-synchrony assumption failed.

\paragraph{Case III: Expanding volume with spatial scaling.}
We define $\tau = N\tau_0$ and $c = 1$, yielding $M = N-1$ and $\Omega = \langle \omega \rangle / (1 + (N-1)K\tau_0)$. As $N$ grows, delays increase, eventually de-synchronizing the network and invalidating the linear approximation (Fig.~\ref{fig:fig1}c).

\paragraph{Case IV: Increasing density in fixed volume.}
Delays scale inversely with $\sqrt{N}$: $\tau = \tau_0/\sqrt{N}$ as in a bi-dimensional space the distances increase as $N^{1/2}$ \cite{Albert2002}, and weights scale with an inverse square rule found in multiple physical phenomena as $c = c_0/\tau^2 = N c_0/\tau_0^2$. This yields $M = (N-1)Nc_0/\tau_0^2$ and the expression $\Omega = \langle \omega \rangle / (1 + (N-1)Kc_0/(\sqrt{N}\tau_0))$. Simulations confirm the accuracy of the analytic prediction across $N$ with full agreement (Fig.~\ref{fig:fig1}d).

\paragraph{Case A: Expanding the radius of a fully-connected circular network.}
 First, we used fully-connected networks where the radius increases as $r=N\tau_0$ and the weights are determined by Eq~\eqref{eq:heterogenous_weights}. The nodes are located in the circumference of radius $r$ (Fig.~\ref{fig:fig2}a). The collective frequency decreases following the analytical prediction.

 \paragraph{Case A: Expanding the radius of a fixed-degree circular network.}
 Second, we used fixed-degree networks using the $k=4$ farther nodes (Fig.~\ref{fig:fig2}b) and the $k=4$ nearer nodes (Fig.~\ref{fig:fig2}c). In both cases, we did not achieved synchrony for $N>5$ ($\langle KOP \rangle\ \approx 0  $), thus the collective frequency does not decreases, and interestingly the prediction for the nearer nodes is an increase in the collective frequency.   

 \paragraph{Case A: Expanding the radius of a fixed-density circular network.}
 Third, we used fixed density networks using $p=0.6$ (Fig.~\ref{fig:fig2}d), and $p=0.2$ (Fig.~\ref{fig:fig2}e) keeping the connections with the nearer nodes. When synchrony is achieved ($\langle KOP \rangle\ \approx 1  $), the collective frequency decreases. In contrast to fully-connected networks, the variance of the intrinsic frequencies are relevant, then the numerical results varies for each initial condition.

\paragraph{Case A: Expanding the radius of a fixed-density circular network with random connections.}
 Finally, we used fixed density networks using $p=0.6$ (Fig.~\ref{fig:fig2}f), and $p=0.2$ (Fig.~\ref{fig:fig2}g) but allowing random non-symmetric connections. Synchrony is achieved ($\langle KOP \rangle\ \approx 1 $) using p=0.6, and numerical results are close to the analytical predictions. For $p=0.2$, synchrony is not achieved and the numerical collective frequency remains close to the average intrinsic frequency (40 Hz).

\paragraph{Case B: Increasing spatial density in fixed radius circular network.}
In this last case, the radius of the circle keeps constant while more nodes are located in the circumference. The delays decay asymptotically while weights increases proportionally with N. Synchrony is mostly achieved for all the number of nodes, showing a small variance above the predicted collective frequency for the network's sizes where $\langle KOP \rangle$ is near 1. 


\section{Discussion}

We analytically derived how the collective frequency of a network of coupled phase oscillators depends on network size under distinct growth scenarios. These scenarios capture realistic constraints found in biological and artificial systems: fixed or expanding spatial volume, increasing node density, and scaling of delays and weights. By assuming steady state phase synchrony, we arrived at closed-form expressions that reveal how collective frequency shifts as a function of delay, coupling strength, and network structure. The derived collective frequency is also the intrinsic resonance frequency as it comes from the system response at the intrinsic frequencies of the individual oscillators. 

In all realizable cases—i.e., those incorporating geometry or growth—collective frequency decreases as the number of nodes increases. This decay arises from the compounded effect of interaction delays and cumulative connectivity. While in the idealized case of a non-geometric network the collective frequency approaches the mean intrinsic frequency~\cite{Strogatz2000}, incorporating delays—especially increasing delays—leads to systematic slowing of global dynamics. These analytical predictions match well with numerical simulations, provided the system remains phase-synchronized.

One key insight is that the topology and physical embedding of the network shape how delays accumulate. In Case IV (fixed volume), denser parcellation leads to stronger cumulative interactions, whereas in Case III (expanding volume), longer distances increase delays, further reducing the collective frequency. In both scenarios, synchronization may break down for large $N$, as indicated by the decline of the Kuramoto Order Parameter. This breakdown marks the regime where our steady-state approximations no longer hold.

The Case III was further analyzed with the real spatial values in case A by varying the topology of the expanding network as fully-connected, fixed degree or fixed connectivity density. In the last case using connections between the nearer nodes but also with random connections. As expected, the networks with lower number connections and larger delays even fail to synchronize using the same global coupling that synchronizes the $N=5$ network. However, if the density of the network is enough to achieve synchronization, then the numerical results follow the analytical results. Achieving this prediction have importance for real-world problems where the networks are weighted and directed.   

From Case A results, we suggest the average intrinsic frequency as a suitable guess for the collective frequency when the system does not achieve synchronization. However, the results also suggest that at intermediate levels of synchrony the collective frequency decreases if the heavier connections are a) enough density of the network and b) the ones with the increasing delays. 

Our results provide a rigorous framework for understanding how large-scale network properties shape collective timescales. These findings have implications in neuroscience, where synchronized oscillations govern cognition, memory, and perception. For example, the slowing of alpha and beta rhythms in aging or neurodegeneration could partly reflect increased effective delays due to white matter degradation or network reconfiguration. Moreover, brain network models using parcellation-based simulations (e.g., via the Kuramoto model) must account for how node resolution and spatial embedding nonlinearly affect emergent dynamics.

From a theoretical standpoint, the analysis complements prior works on synchronization in complex networks~\cite{Arenas2008, Lee2009} and extends the results of Niebur et al.~\cite{Niebur1991} by considering scaling properties and topological embeddings. The introduction of heterogeneous delays further aligns our work with more realistic models of brain networks.

Although we refer to cortical dynamics as a familiar example of a spatially extended oscillator network, the mechanism derived here is not specific to neuroscience. Any delayed-coupled system with finite conduction speed exhibits the same size-dependent resonance predicted by the analytical solution, with the brain cortex simply providing a well-characterised empirical instance.

\paragraph{Limitations and Future Directions.} Our analysis relies on the near-synchrony approximation—a standard but necessary simplification for obtaining closed-form solutions in coupled oscillator systems with heterogeneous delays. While this assumption may not hold in weakly connected, highly heterogeneous, or biologically realistic networks, it provides a tractable foundation to reveal how structural factors shape emergent dynamics. Notably, regimes with increasing delays or high density can exhibit a breakdown of global synchrony, giving rise to metastable dynamics characterized by transient, localized coherence. In such cases, the predicted collective frequency still captures the dominant organizing timescale, around which the system self-organizes before transitioning between states. This suggests that the derived scaling laws remain informative beyond perfect synchrony and offer insight into the structural constraints of multistate behavior. Future work should extend the framework to include noise, hierarchical and modular topologies, and biologically plausible plasticity mechanisms—such as frequency or spike-timing-dependent adaptations—to better capture the evolving dynamics of real-world networks.

\section{Data Availability}
The code needed to reproduce the simulations and figures is open available \cite{github}.

\section{Acknowledgments}
The authors are grateful to Professor Peter Robinson for his insightful comments on an earlier version of the manuscript. This work is supported by ANID, Chile, projects Exploracion 13240064; FONDECYT  1241695 (WED); BASAL AFB240002 (WED and AW); and 3230682 (FT). JMC acknowledges financial support from Ikerbasque: The Basque Foundation for Science, and from Spanish Ministry of Science (PID2023-148012OB-I00), Spanish Ministry of Health (PI22/01118), Basque Ministry of Health (2023111002 \& 2022111031).

\appendix
\section{Growing Circular Network}

\begin{figure*}[!ht]
    \centering
    \includegraphics[width=\linewidth]{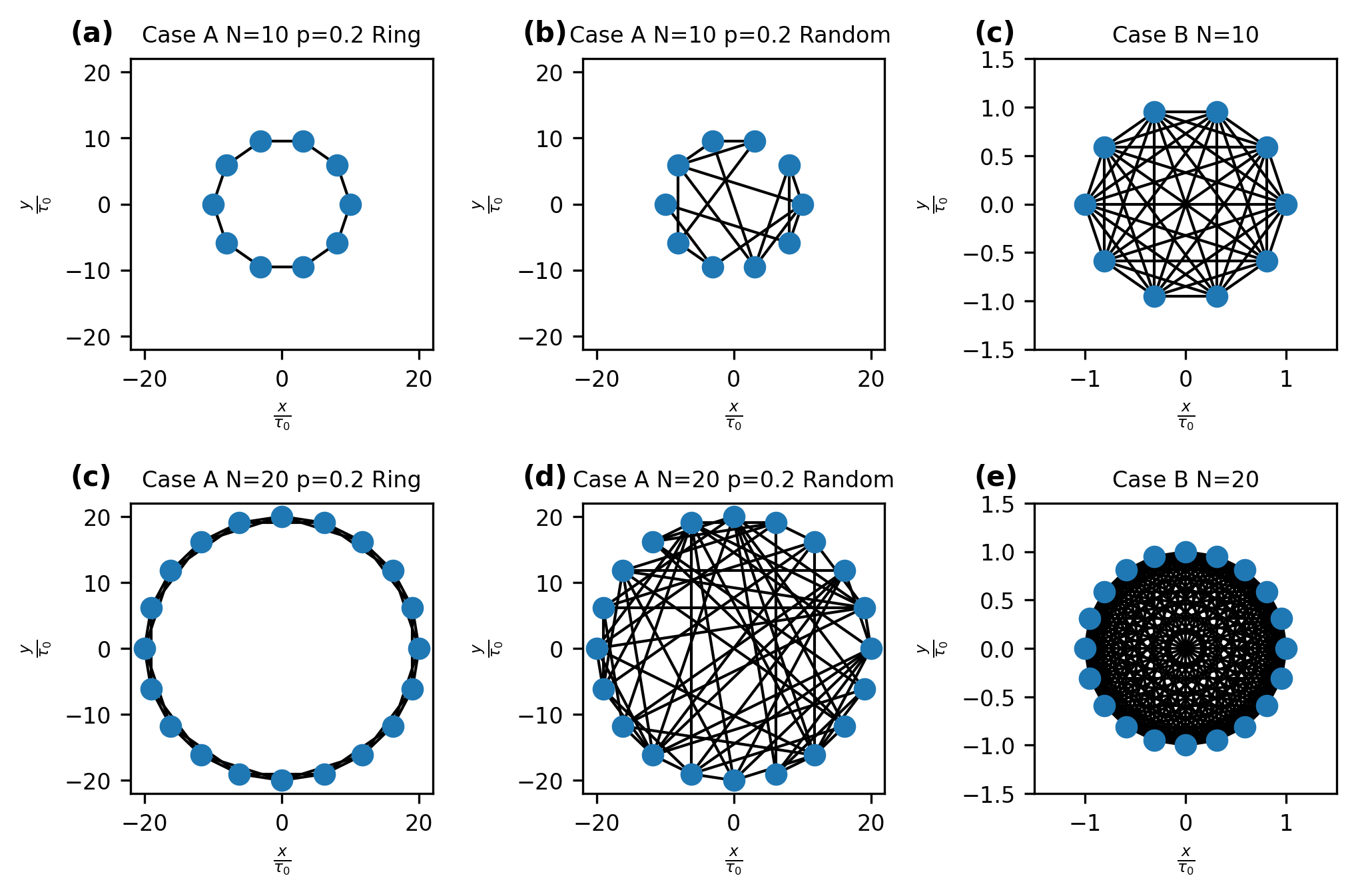}
    \caption{{\bf Examples of the circular network}. 
    (a) Case A network for $N=10$ with connectivity density of $p=0.2$ keeping the nearer connections. 
    (b) Case A network for $N=10$ with connectivity density of $p=0.2$ using random connections.
    (d) Case A network for $N=20$ with connectivity density of $p=0.2$ keeping the nearer connections.
    (e) Case A network for $N=20$ with connectivity density of $p=0.2$ using random connections.
    (c) Case B network for $N=10$ fully-connected.
    (f) Case B network for $N=20$ fully-connected.}
    \label{fig:fig_a}
\end{figure*}

The network used in cases A and B have located its nodes in the circumference of a circle centered in the origin. In case A, the radius of the circle increases along with the number of nodes. There are several variations of the density (sparsity) of the connectivity between the nodes. Fig.~\ref{fig:fig_a} (a)-(b) show the network for $N=10$ and Fig.~\ref{fig:fig_a} (d)-(e) the netwrok when the number of nodes increased to $N=20$. In case B, the radius does not vary, only more nodes are added to the circumference. For this cases we only used fully-connected networks. Fig.~\ref{fig:fig_a} (c) shows the network for $N=10$ and Fig.~\ref{fig:fig_a} (f) shows the network for $N=20$ in case B.

\bibliography{apssamp}

@article{Strogatz2000,
  title={From Kuramoto to Crawford: exploring the onset of synchronization in populations of coupled oscillators},
  author={Strogatz, Steven H},
  journal={Physica D: Nonlinear Phenomena},
  volume={143},
  number={1-4},
  pages={1--20},
  year={2000},
  publisher={Elsevier}
}

@article{Breakspear2010,
  title={Generative models of cortical oscillations: neurobiological implications of the Kuramoto model},
  author={Breakspear, Michael and Heitmann, Stewart and Daffertshofer, Andreas},
  journal={Frontiers in human neuroscience},
  volume={4},
  pages={190},
  year={2010},
  publisher={Frontiers Research Foundation}
}

@article{Niebur1991,
  title={Collective frequencies and metastability in networks of limit-cycle oscillators with time delay},
  author={Niebur, Ernst and Schuster, Heinz G and Kammen, Daniel M},
  journal={Physical review letters},
  volume={67},
  number={20},
  pages={2753},
  year={1991},
  publisher={APS}
}

@article{Cabral2014,
  title={Exploring mechanisms of spontaneous functional connectivity in MEG: how delayed network interactions lead to structured amplitude envelopes of band-pass filtered oscillations},
  author={Cabral, Joana and Luckhoo, Henry and Woolrich, Mark and Joensson, Morten and Mohseni, Hamid and Baker, Adam and Kringelbach, Morten L and Deco, Gustavo},
  journal={Neuroimage},
  volume={90},
  pages={423--435},
  year={2014},
  publisher={Elsevier}
}

@inproceedings{Kuramoto1975,
  title={Self-entrainment of a population of coupled non-linear oscillators},
  author={Kuramoto, Yoshiki},
  booktitle={International Symposium on Mathematical Problems in Theoretical Physics: January 23--29, 1975, Kyoto University, Kyoto/Japan},
  pages={420--422},
  year={1975},
  organization={Springer}
}

@article{Lee2009,
  title={Large coupled oscillator systems with heterogeneous interaction delays},
  author={Lee, Wai Shing and Ott, Edward and Antonsen, Thomas M},
  journal={Physical review letters},
  volume={103},
  number={4},
  pages={044101},
  year={2009},
  publisher={APS}
}

@article{Torres2024,
title={Emergence of multiple spontaneous coherent subnetworks from a single configuration of human connectome coupled oscillators model},
  author={Torres, Felipe A and Otero, M{\'o}nica and Lea-Carnall, Caroline A and Cabral, Joana and Weinstein, Alejandro and El-Deredy, Wael},
  journal={Scientific Reports},
  volume={14},
  number={1},
  pages={30726},
  year={2024},
  publisher={Nature Publishing Group UK London}
}

@article{LeaCarnall2016,
   author = {Caroline A. Lea-Carnall and Marcelo A. Montemurro and Nelson J. Trujillo-Barreto and Laura M. Parkes and Wael El-Deredy},
   doi = {10.1371/journal.pcbi.1004740},
   issn = {15537358},
   issue = {2},
   journal = {PLoS Computational Biology},
   pages = {1-19},
   pmid = {26914905},
   title = {Cortical Resonance Frequencies Emerge from Network Size and Connectivity},
   volume = {12},
   year = {2016},
}

@article{Arenas2008,
  title={Synchronization in complex networks},
  author={Arenas, Alex and D\'{i}az-Guilera, Albert and Kurths, Jurgen and Moreno, Yamir and Zhou, Changsong},
  journal={Physics reports},
  volume={469},
  number={3},
  pages={93--153},
  year={2008},
  publisher={Elsevier},
}

@article{vanWijk2010,
  title={Comparing brain networks of different size and connectivity density using graph theory},
  author={Van Wijk, Bernadette CM and Stam, Cornelis J and Daffertshofer, Andreas},
  journal={PloS one},
  volume={5},
  number={10},
  pages={e13701},
  year={2010},
  publisher={Public Library of Science San Francisco, USA}
}

@article{salova2025,
  title={Combined topological and spatial constraints are required to capture the structure of neural connectomes},
  author={Salova, Anastasiya and Kov{\'a}cs, Istv{\'a}n A},
  journal={Network Neuroscience},
  volume={9},
  number={1},
  pages={181--206},
  year={2025},
  publisher={MIT Press 255 Main Street, 9th Floor, Cambridge, Massachusetts 02142, USA~…}
}

@article{oDea2013,
  title={Spreading dynamics on spatially constrained complex brain networks},
  author={O'Dea, Reuben and Crofts, Jonathan J and Kaiser, Marcus},
  journal={Journal of the Royal Society Interface},
  volume={10},
  number={81},
  pages={20130016},
  year={2013},
  publisher={The Royal Society}
}

@article{boccaletti2006,
  title={Complex networks: Structure and dynamics},
  author={Boccaletti, Stefano and Latora, Vito and Moreno, Yamir and Chavez, Martin and Hwang, D-U},
  journal={Physics reports},
  volume={424},
  number={4-5},
  pages={175--308},
  year={2006},
  publisher={Elsevier}
}

@article{Pang2023,
  title={Geometric constraints on human brain function},
  author={Pang, James C and Aquino, Kevin M and Oldehinkel, Marianne and Robinson, Peter A and Fulcher, Ben D and Breakspear, Michael and Fornito, Alex},
  journal={Nature},
  volume={618},
  number={7965},
  pages={566--574},
  year={2023},
  publisher={Nature Publishing Group UK London}
}

@article{Oliveira2020,
  title={Nonmonotonic critical threshold in the kuramoto model},
  author={de Oliveira, JF and Abud, CV},
  journal={Communications in Nonlinear Science and Numerical Simulation},
  volume={91},
  pages={105428},
  year={2020},
  publisher={Elsevier}
}

@article{Albert2002,
  title = {Statistical mechanics of complex networks},
  author = {Albert, R\'eka and Barab\'asi, Albert-L\'aszl\'o},
  journal = {Rev. Mod. Phys.},
  volume = {74},
  issue = {1},
  pages = {47--97},
  numpages = {0},
  year = {2002},
  month = {Jan},
  publisher = {American Physical Society},
  doi = {10.1103/RevModPhys.74.47},
  url = {https://link.aps.org/doi/10.1103/RevModPhys.74.47}
}

@article{Coombes2010,
  title={Large-scale neural dynamics: simple and complex},
  author={Coombes, Stephen},
  journal={NeuroImage},
  volume={52},
  number={3},
  pages={731--739},
  year={2010},
  publisher={Elsevier}
}

@article{iniguez2023universal,
  title={Universal patterns in egocentric communication networks},
  author={I{\~n}iguez, Gerardo and Heydari, Sara and Kert{\'e}sz, J{\'a}nos and Saram{\"a}ki, Jari},
  journal={Nature Communications},
  volume={14},
  number={1},
  pages={5217},
  year={2023},
  publisher={Nature Publishing Group UK London}
}

@article{Daqing2011,
  title={Dimension of spatially embedded networks},
  author={Daqing, Li and Kosmidis, Kosmas and Bunde, Armin and Havlin, Shlomo},
  journal={Nature Physics},
  volume={7},
  number={6},
  pages={481--484},
  year={2011},
  publisher={Nature Publishing Group UK London}
}

@misc{knp,
title={https://github.com/AnilloViBrain/KuramotoNetworksPackage},
author={BrainDynamicsLaboratory},
year={2024},
howpublished={GitHub repository},
url={https://github.com/AnilloViBrain/KuramotoNetworksPackage}
}

@misc{github,
title={https://github.com/AnilloViBrain/NetworksSizeResonance},
author={BrainDynamic Laboratory},
year={2025},
howpublished={GitHub repository},
url={https://github.com/AnilloViBrain/NetworksSizeResonance}
}

\end{document}